\documentstyle[12pt]{article}

\newfont{\bg}{cmr10 scaled\magstep4}

\newcounter{figurenumber}
\renewcommand{\caption}[2]{%
\par\begingroup\noindent\refstepcounter{figurenumber}%
{\bf Fig.\ \arabic{figurenumber}:} \label{#1} #2
\par\endgroup}
\catcode `\@=11
\def\section{\@startsection {section}{1}{\z@}{-3.5ex plus -1ex minus
 -.2ex}{2.3ex plus .2ex}{\Large\bf}}
\def\subsection{\@startsection{subsection}{2}{\z@}{-3.25ex plus -1ex minus
 -.2ex}{1.5ex plus .2ex}{\large}}
\def\subsubsection{\@startsection{subsubsection}{3}{\z@}{-3.25ex plus
 -1ex minus -.2ex}{1.5ex plus .2ex}{\normalsize\it}}
\catcode `\@=12

\begin{document}
\begin{center}
{\Large\bf
  Scaling Theory of Antiferromagnetic\\
  Heisenberg Ladder Models \par}
\vskip 1.5em
{\large \lineskip .5em
\begin{tabular}[t]{c}
  Naomichi Hatano{\footnotemark[1]}\ and Yoshihiro Nishiyama\\
  {\normalsize\it
    Department of Physics, University of Tokyo,
    Hongo 7-3-1, Bunkyo-ku, Tokyo 113, Japan}
\end{tabular}
\par}
\end{center}
\par

\footnotetext[1]{Address from April 1995 to March 1996:
Lyman Laboratory of Physics, Harvard University, Cambridge, Massachusetts
02138, USA}

\begin{center}{\bf Abstract\vspace{-.5em}\vspace{0pt}}\end{center}
\begin{quotation}
The $S=1/2$ antiferromagnetic Heisenberg model on multi-leg ladders is
investigated.
Criticality of the ground-state transition is explored by means of finite-size
scaling.
The ladders with an even number of legs and those with an odd number of legs
are distinguished clearly.
In the former, the energy gap opens up as $\Delta E\sim{J_\perp}$, where
${J_\perp}$ is the strength of the antiferromagnetic inter-chain coupling.
In the latter, the critical phase with the central charge $c=1$ extends over
the whole region of ${J_\perp}>0$.
\end{quotation}

\section{Introduction}
\setcounter{equation}{0}
\label{intro}
Understanding of the ground-state criticality of one-dimensional quantum
system has been enriched greatly in the past decade.
One of the important breakthroughs was finite-size scaling based on
the conformal field theory [\ref{Cardy84a}];
see [\ref{Cardy87}] for a review.
Nowadays many attempts on generalization for higher-dimensional systems are in
progress.
One of them is the study of coupled chains.
The Heisenberg ladder models are particularly of interest.
The two-dimensional Heisenberg model might be explored as the limiting case of
the ladder models.
In addition, actual substances which realize the ladder models have been
developed experimentally: Sr$_{2n-2}$Cu$_{2n}$O$_{4n-2}$, for example
[\ref{Takano92}].

The Hamiltonian of the $S=1/2$ Heisenberg ladder models is given by
\begin{equation}\label{1-10}
{\cal H}\equiv{{\cal H}_{\rm leg}}+{{\cal H}_{\rm rung}},
\end{equation}
where
\begin{eqnarray}\label{1-20}
{{\cal H}_{\rm leg}}&\equiv&J\sum_{x=0}^{L-1}\sum_{y=1}^{{n_l}}
\vec{S}_{x,y}\cdot\vec{S}_{x+1,y},
\\
\label{1-30}
{{\cal H}_{\rm rung}}&\equiv&{J_\perp}\sum_{x=0}^{L-1}\sum_{y=1}^{{n_l}-1}
\vec{S}_{x,y}\cdot\vec{S}_{x,y+1}.
\end{eqnarray}
In the present paper we treat the antiferromagnetic case,
$J>0$ and ${J_\perp}>0$.
Each component of the spins is defined by
\begin{equation}\label{1-40}
S^\mu=\frac{1}{2}\sigma^\mu
\qquad
(\mu=x,y,z),
\end{equation}
where $\{\sigma^\mu\}$ are the Pauli matrices.
We define $S^\pm\equiv S^x\pm{\rm i}S^y$.
We impose periodic boundary conditions only in the $x$ direction:
\begin{equation}\label{1-50}
\vec{S}_{L,y}=\vec{S}_{0,y}
\qquad\mbox{for}\quad
y=1,2,\ldots,{n_l}.
\end{equation}
Hereafter the system size $L$ is even.

In the present paper, we focus on the ground-state criticality of the ladder
models.
It was conjectured [\ref{White94}] that the ladder models with even ${n_l}$
and those with odd ${n_l}$ behave quite differently from each other;
for ${J_\perp}=1$, the former is massive, while the latter is massless.
Here we discuss the criticality for arbitrary positive ${J_\perp}$ and for
general ${n_l}$.

It is well known that the antiferromagnetic Heisenberg chain (${J_\perp}/J=0$)
is massless [\ref{Bethe31}].
Many questions arise when we introduce the inter-chain coupling
${{\cal H}_{\rm rung}}$:
It has been quite controversial [\ref{Dagotto92}-\ref{Terai95}]
whether the system with ${n_l}=2$ becomes
massive for any positive ${J_\perp}$, or there exists a critical point at a
finite value of ${J_\perp}$;
It has not been clarified whether the massless phase for odd ${n_l}$
extends over the whole region of ${J_\perp}>0$, or a massive phase exists in
the region $0<{J_\perp}<1$;
The universality of the massless phase has not been discussed either.

The purpose of the present paper is to answer the above questions on the basis
of finite-size scaling theory.
We introduce our scaling ansatz in section \ref{sec:ansatz}.
The ansatz yields useful conclusions immediately.
In particular, we conclude that the whole region of ${J_\perp}>0$ is controlled
by the stable fixed point at ${J_\perp}/J=\infty$.
Thus we distinguish the systems with even ${n_l}$ and those with  odd ${n_l}$
clearly:
the former is massive for any ${J_\perp}$,
while the latter is critical with the central charge $c=1$ for any ${J_\perp}$.
We present some results for even ${n_l}$ and for odd ${n_l}$ in sections
\ref{sec:even} and \ref{sec:odd}, respectively.
We particularly show estimation of critical amplitudes.
We give numerical confirmation of the scaling ansatz in section
\ref{sec:ansatz}.
In addition to that, we confirm the ansatz by means of
perturbation theory in section \ref{sec:pert}.

\section{Finite-size scaling of the energy gap}
\setcounter{equation}{0}
\label{sec:scaling}
In this section, we introduce the finite-size scaling form of the energy gap.
We numerically confirm the scaling ansatz.
We present our conclusions drawn from the ansatz.

\subsection{General arguments}
\label{sec:ansatz}
The central scaling ansatz of the present paper is written in the form
\begin{equation}\label{2-10}
\Delta E(L)\simeq \frac{J}{L}{\tilde{\Delta}}(x)
={J_\perp}\frac{{\tilde{\Delta}}(x)}{x}
\qquad\mbox{with}\quad
x\equiv\frac{L{J_\perp}}{J},
\end{equation}
where $\Delta E$ is the energy gap between the ground state and the first
excited state, and ${\tilde{\Delta}}$ denotes the relevant scaling function.
We show, in figure \ref{scale},
numerical data obtained by the Lanczos method
[\ref{Lanczos50}].
\begin{figure}[tbh]
\begin{minipage}[t]{182mm}
\special{epsfile=fig1a_pre.eps hsize=255 voffset=14}
\raisebox{-60mm}{(a)}
\hspace*{89mm}
\special{epsfile=fig1b_pre.eps hsize=255 voffset=14}
\raisebox{-60mm}{(b)}
\vspace*{10mm}
\end{minipage}
\\
\begin{minipage}[t]{98mm}
\special{epsfile=fig1c_pre.eps hsize=255 voffset=14}
\raisebox{-60mm}{(c)}
\hspace*{97mm}
\end{minipage}
\begin{minipage}[t]{83mm}
\vspace*{5mm}
\caption{scale}%
{Scaling plot of numerical data for the ${J_\perp}$ dependence of the energy
gap.
The abscissa is scaled as $x\equiv L{J_\perp}/J$, while the ordinate is scaled
as $L\Delta E/J$:
(a) ${n_{\rm leg}}=2$, $4\leq L\leq12$; (b) ${n_{\rm leg}}=3$, $4\leq L\leq8$;
(c) ${n_{\rm leg}}=4$, $4\leq L\leq6$.}
\end{minipage}
\end{figure}
The data are scaled well over a finite region of $x$ (namely $0\leq x\leq5$)
for ${n_l}=2,4$, and over the whole region of $x$ for ${n_l}=3$.
We confirm the scaling form (\ref{2-10}) by means of perturbation theory in
section \ref{sec:pert}.

The ansatz (\ref{2-10}) implies that the inter-chain coupling is a relevant
operator.
The coupling ${{\cal H}_{\rm rung}}$ drives the system away from the point
${J_\perp}/J=0$, and makes the system renormalized to the limit
${J_\perp}/J\to\infty$ as $L\to\infty$; see figure \ref{renom}.
\begin{figure}[t]
\begin{minipage}[t]{99mm}
\special{epsfile=fig2_pre.eps hsize=255}
\hspace*{99mm}\vspace*{45mm}
\end{minipage}
\begin{minipage}[t]{83mm}
\caption{renom}%
{The conjectured renormalization-flow diagram for the Heisenberg ladders.
The fixed point ${J_\perp}/J=0$ is unstable with respect to the inter-chain
coupling.
The stable fixed point at ${J_\perp}/J=\infty$ is described by the massive
$S=0$ chain for even ${n_{\rm leg}}$, and by the massless $S=1/2$ chain for
odd ${n_{\rm leg}}$.
We thereby conclude that the whole phase of ${J_\perp}>0$ is massive for even
${n_{\rm leg}}$ and is critical for odd ${n_{\rm leg}}$.}
\end{minipage}
\end{figure}
Thus the fixed point at ${J_\perp}/J=\infty$ controls the whole phase of the
region ${J_\perp}>0$.
Apart from correction to scaling, the thermodynamic limit $L\to\infty$  is
equivalent to the limit ${J_\perp}/J\to\infty$.

We can thereby understand the difference between the systems with even ${n_l}$
and those with odd ${n_l}$.
For even ${n_l}$, the spins on each rung form an $S=0$ singlet in the limit
${J_\perp}/J\to\infty$.
The system at the fixed point ${J_\perp}/J=\infty$ is the $S=0$ chain
with the energy gap of the order of ${J_\perp}$ [\ref{Reigrotzki95}].
For odd ${n_l}$, on the other hand, the spins on each rung form an $S=1/2$
doublet in the limit ${J_\perp}/J\to\infty$.
In the first-order perturbation of $J$, we have the same energy spectrum as of
the $S=1/2$ Heisenberg chain.
(See section \ref{sec:odd} for details.)
Hence the system at the fixed point ${J_\perp}/J=\infty$ is the $S=1/2$
Heisenberg chain, which is massless, or critical.
Thus the whole phase of ${J_\perp}>0$ for even (odd) ${n_l}$ is controlled by
the massive $S=0$ (massless $S=1/2$) Heisenberg chain.

In order to argue asymptotic behavior of the scaling function
${\tilde{\Delta}}(x)$, we can exploit the two limits ${J_\perp}/J\to\infty$ and
$L\to\infty$.
In the thermodynamic limit $L\to\infty$, we must have non-divergent value of
the energy gap.
Hence the scaling function ${\tilde{\Delta}}(x)$ in (\ref{2-10}) must be of the
first or lower order of  $x$ in the limit $x\to\infty$.
We then postulate that the asymptotic form
differs for even ${n_l}$ and for odd ${n_l}$ as follows:
\begin{equation}\label{2-20}
{\tilde{\Delta}}(x)\sim
\left\{\begin{array}{ll}
Ax&\quad\mbox{for even ${n_l}$}\\
b&\quad\mbox{for odd ${n_l}$}
\end{array}\right.
\qquad\mbox{as}\quad
x\to\infty.
\end{equation}
Here $A$ and $b$ are appropriate constants.
We can derive this from the behavior of the energy gap in the limit
${J_\perp}/J\to\infty$.
For even ${n_l}$, a first excited state is created by exciting one of the
$S=0$ singlets on the rungs.
The energy gap to this state is proportional to ${J_\perp}$:
$\Delta E\simeq{J_\perp}{\tilde{\Delta}}(x)/x\propto{J_\perp}$.
Therefore we should have ${\tilde{\Delta}}(x)\sim x$.
For odd ${n_l}$, the energy gap for the first excited state is that of the
$S=1/2$ Heisenberg chain, and hence is proportional to $J/L$:
$\Delta E\simeq(J/L){\tilde{\Delta}}(x)\propto J/L$.
This yields ${\tilde{\Delta}}(x)\sim \mbox{constant}$.
The above postulate (\ref{2-20}) is actually observed in the numerical data in
figure \ref{scale}.

\subsection{Ladders with an even number of legs}
\label{sec:even}
According to the above ansatz (\ref{2-20}), we have
${\tilde{\Delta}}(x)/x\to A$ as $x\to\infty$ for even ${n_l}$.
Hence in the thermodynamic limit $L\to\infty$, the scaling form (\ref{2-10}) is
reduced to
\begin{equation}\label{2-30}
\Delta E\simeq A{J_\perp}^\nu
\qquad\mbox{with}\quad
\nu=1.
\end{equation}
The coefficient $A$ defined in (\ref{2-20}) turns out to be the critical
amplitude.
Thus the critical system at ${J_\perp}/J=0$ (namely, a set of the independent
Heisenberg  chains) becomes massive immediately when we turn on the inter-chain
coupling.
The critical-point estimate ${J_\perp}/J=0$ and the exponent estimate $\nu=1$
were concluded previously for $n_l=2$ [\ref{Nishiyama95b}, \ref{Terai95}].

Let us estimate the critical amplitude $A$ for ${n_l}=2$.
We utilize the following general finite-size scaling form for one-dimensional
quantum ground-state phase transitions [\ref{Hatano94c}]:
\begin{equation}\label{2-40}
\frac{L\Delta E(L)}{J}\simeq
\left\{\begin{array}{ll}
Ax+D_+{\rm e}^{-Cx} & \quad\mbox{in the disordered phase},\\
B & \quad\mbox{at the critical point},\\
D_-{\rm e}^{-Cx} & \quad\mbox{in the ordered phase}.
\end{array}\right.
\end{equation}
Here $x\equiv L|\varepsilon|^\nu$ denotes the relevant scaling variable with
$\varepsilon$ being the distance from the critical point;
$x=L{J_\perp}/J$ in the present case.
The coefficient $A$ gives the critical amplitude in (\ref{2-30}).
The coefficient $C$, on the other hand, gives the critical amplitude of the
correlation length: $\xi^{-1}\simeq C({J_\perp}/J)^{\nu}$.
We introduced in [\ref{Hatano94c}] the amplitude relation
\begin{equation}\label{2-60}
A/C={v_{\rm s}},
\end{equation}
where ${v_{\rm s}}$ is the sound velocity.
The coefficient $B$ is predicted from the conformal field theory as
$B={v_{\rm s}}\pi\eta$, where $\eta$ is Fisher's correlation exponent
[\ref{Fisher64b}].
For the Hamiltonian (\ref{1-10}) with ${J_\perp}/J=0$, the Bethe-ansatz
solution [\ref{Alcaraz88a}] gives
\begin{equation}\label{2-50}
\eta=1
\quad\mbox{and}\quad
{v_{\rm s}}=\frac{\pi}{2};
\end{equation}
hence we have
\begin{equation}\label{2-55}
\frac{A}{C}=\frac{\pi}{2}=1.570796\cdots
\qquad\mbox{and}\qquad
B=\frac{\pi^2}{2}=4.934802\cdots.
\end{equation}
The coefficients $D_+$ and $D_-$ may have weak $x$ dependence
[\ref{Hatano94c}].

The asymptotic behavior in (\ref{2-20}) for even ${n_l}$ is consistent with the
scaling form (\ref{2-40}) in the disordered phase.
We can employ our previous analysis in [\ref{Hatano94c}], where we
showed for the one-dimensional quantum Potts model that a good estimator of the
amplitude $A$ is given by
\begin{equation}\label{2-70}
A_L^{\rm min}\equiv
\min_{x>0}\left(\frac{L\Delta E/J}{x}\right).
\end{equation}
(If the data followed the scaling form (\ref{2-40}) completely, the quantity
$L\Delta E/(Jx)$ would converge to $A$ exponentially as $x\to\infty$.
However, there appears the minimum in practice, probably because some
correction to scaling [\ref{Hatano94c}].)
The estimator (\ref{2-70}) may converge to $A$ exponentially [\ref{Hatano94c}]
in the form
\begin{equation}\label{2-80}
A_L^{\rm min}\simeq A+c_1\exp(-c_2L).
\end{equation}
We calculated $A_L^{\rm min}$ for the data in figure \ref{scale} (a), and
fitted the results to the form (\ref{2-80});
see figure \ref{fit}.
\begin{figure}[t]
\begin{minipage}[t]{182mm}
\special{epsfile=fig3a_pre.eps hsize=255 voffset=14}
\raisebox{-60mm}{(a)}
\hspace*{89mm}
\special{epsfile=fig3b_pre.eps hsize=255 voffset=14}
\raisebox{-60mm}{(b)}
\vspace*{5mm}
\end{minipage}
\caption{fit}%
{Estimation of the critical amplitude $A$ for ${n_{\rm leg}}=2$ as was done in
[\ref{Hatano94c}].
(a) The quantity $L\Delta E/(Jx)$ versus $x$ for $L=12$, for example.
The estimator (\ref{2-70}) is obtained as
$A_{12}^{\rm min}=0.505\cdots$, which is indicated by the dotted lines.
(b) The values of the estimator (\ref{2-70}) are fitted to the form
(\ref{2-80}).
The solid line shows the best fit
$A_L^{\rm min}\simeq0.468+0.505\exp(-0.218L)$.}
\end{figure}
We thereby have the estimate $A=0.47(1)$ and hence $1/C={v_{\rm s}}/A=3.34(7)$
for ${n_l}=2$.
(The error of the estimate $A$ was evaluated through the least-squares fitting
to the form (\ref{2-80}).)
Namely, we have
\begin{equation}\label{2-90}
\Delta E\simeq0.47{J_\perp}
\qquad\mbox{and}\qquad
\xi\simeq3.34\frac{J}{{J_\perp}}
\end{equation}
as $L\to\infty$ and near ${J_\perp}/J\sim0$.
These explain the following estimates in [\ref{White94}] quite well:
$\Delta E=0.504J$ and $\xi=3.19(1)$ for ${n_l}=2$ and ${J_\perp}/J=1$.
Although it is unnecessary that the critical behavior (\ref{2-90}) near
${J_\perp}/J\sim0$ is observed even for ${J_\perp}/J=1$,  we see in figure 2 of
[\ref{Barnes93}] that correction to (\ref{2-90}) may be quite small in the
region ${J_\perp}/J\leq1$.

Incidentally, the energy gap is $\Delta E={J_\perp}$ for ${n_l}=2$ in the limit
${J_\perp}/J\to\infty$,  where the system is reduced to a set of independent
dimers.
A crossover from the behavior (\ref{2-90}) to the behavior $\Delta E={J_\perp}$
may occur in the region ${J_\perp}/J>1$.

If we naively apply the same analysis as above to the data for ${n_l}=4$ with
$L=2$, $4$ and $6$, we obtain the tentative estimates
\begin{equation}\label{2-95}
A\simeq0.27
\qquad\mbox{and}\qquad
C^{-1}={v_{\rm s}}/A\simeq5.8.
\end{equation}
Though there may be some errors in these estimates,
the values (\ref{2-90}) and (\ref{2-95}) are fairly consistent with the scaling
hypothesis $\xi\propto{n_l}$ [\ref{White94}].

\subsection{Ladders with an odd number of legs}
\label{sec:odd}
For odd ${n_l}$, we can see in figure \ref{scale}
that the scaling region is quite wide.
We may naturally assume that the scaling form (\ref{2-10}) with (\ref{2-20})
is valid for any value of $L$ and ${J_\perp}/J$.
Thus we have the following remarkable conclusion:
for any positive ${J_\perp}$, the energy gap in the thermodynamic limit
$L\to\infty$ behaves as
\begin{equation}\label{2-190}
\frac{\Delta E}{J}\simeq \frac{b}{L},
\end{equation}
where the amplitude $b$ is the coefficient defined in (\ref{2-20}).
Note that this amplitude is independent of ${J_\perp}$ by definition.
In other words, the energy spectrum in the thermodynamic limit for any
positive ${J_\perp}$ is identical to that for ${J_\perp}/J\to\infty$, or
the spectrum of the $S=1/2$ antiferromagnetic Heisenberg chain.
We thus conclude that the whole phase of ${J_\perp}>0$ for odd ${n_l}$ is a
critical phase with the central charge $c=1$.
Note that the central charge is $c={n_l}$ for ${J_\perp}/J=0$, because we have
${n_l}$ systems of $c=1$ at this point.

We can obtain the value of the amplitude $b$ by considering the limit
${J_\perp}/J\to\infty$.
For this purpose, we first describe how to obtain explicitly the effective
Hamiltonian of the $S=1/2$ Heisenberg chain in the limit
${J_\perp}/J\to\infty$.

In the very limit of ${J_\perp}/J=\infty$, or $J/{J_\perp}=0$, the Hamiltonian
is reduced to ${{\cal H}_{\rm rung}}$;
the rungs are independent of each other.
The ground state of the whole system is the direct product of the ground state
of each rung.
On each rung, an odd number of spins are coupled with the antiferromagnetic
interaction ${J_\perp}$.
Hence two ground states of each rung are degenerate with $S=1/2$ and
$S^z=\pm1/2$.
We express the states on the $x$th rung as $\left|{+_x}\right\rangle$ and
$\left|{-_x}\right\rangle$.
The ground states of the whole system have the $2^L$-fold degeneracy.
This degeneracy is lift up in the first-order perturbation of $J/{J_\perp}$, or
${{\cal H}_{\rm leg}}$.
We calculate the first-order perturbation for degenerate states by writing
down the secular equation:
\begin{equation}\label{2-200}
\left|\hat{H}-\lambda\hat{I}\right|=0.
\end{equation}
Here the matrix $\hat{I}$ denotes the identity operator.
The first-order energy is denoted by $\lambda$.
The matrix $\hat{H}$ is a $2^L\times2^L$ matrix which represents the operator
${{\cal H}_{\rm leg}}$ in the subspace of the degenerate ground states.
For example, one of the diagonal elements of $\hat{H}$ is given by
\begin{eqnarray}\label{2-210}
\lefteqn{
\left\langle{-_1-_2+_3-_4\cdots}\right|
{{\cal H}_{\rm leg}}
\left|{-_1-_2+_3-_4\cdots}\right\rangle
}\nonumber\\
&=&
J\sum_{x=0}^{L-1}\sum_{y=1}^{n_l}
\left\langle{-_1-_2+_3-_4\cdots}\right|
\vec{S}_{x,y}\cdot\vec{S}_{x+1,y}
\left|{-_1-_2+_3-_4\cdots}\right\rangle
\nonumber\\
&=&J\sum_{y=1}^{n_l}
\left(
\left\langle{-_1}\right|S^z_{1,y}\left|{-_1}\right\rangle
\left\langle{-_2}\right|S^z_{2,y}\left|{-_2}\right\rangle
+\left\langle{-_2}\right|S^z_{2,y}\left|{-_2}\right\rangle
\left\langle{+_3}\right|S^z_{3,y}\left|{+_3}\right\rangle
+\cdots
\right),
\end{eqnarray}
and one of the nonzero off-diagonal elements is given by
\begin{equation}\label{2-220}
\left\langle{-_1-_2+_3-_4\cdots}\right|
{{\cal H}_{\rm leg}}
\left|{-_1+_2-_3-_4\cdots}\right\rangle
=\frac{J}{2}\sum_{y=1}^{n_l}
\left\langle{-_2}\right|S^-_{2,y}\left|{+_2}\right\rangle
\left\langle{+_3}\right|S^+_{3,y}\left|{-_3}\right\rangle.
\end{equation}

It is apparent that the secular equation (\ref{2-200}) is equivalent to the
eigenvalue equation for the $S=1/2$ Heisenberg chain:
\begin{equation}\label{2-230}
{\cal H}_{\rm eff}={J_{\rm eff}}
\sum_{x=0}^{L-1}
\vec{S}_x\cdot\vec{S}_{x+1}.
\end{equation}
We can see in (\ref{2-210}) that the effective coupling ${J_{\rm eff}}$ is
given by
\begin{equation}\label{2-240}
{J_{\rm eff}}=J\sum_{y=1}^{n_l}
\left(\left\langle{+}\right|S^z_y\left|{+}\right\rangle\right)^2.
\end{equation}
We show numerical results for ${J_{\rm eff}}$ in
figure \ref{Jeff}.
\begin{figure}[t]
\begin{minipage}[t]{99mm}
\special{epsfile=fig4_pre.eps hsize=255 voffset=25}
\hspace*{99mm}\vspace*{60mm}
\end{minipage}
\begin{minipage}[t]{83mm}
\caption{Jeff}%
{The effective coupling ${J_{\rm eff}}$ for odd ${n_{\rm leg}}$ in the limit
${J_\perp}/J\to\infty$.
The values were calculated numerically on the basis of (\ref{2-240}).}
\end{minipage}
\end{figure}
The value of ${J_{\rm eff}}$ is generally close to unity but increases
monotonically.
In particular, ${J_{\rm eff}}=J$ for ${n_l}=3$ [\ref{Reigrotzki95}].
For large ${n_l}$, the matrix element in (\ref{2-240}) may depend on ${n_l}$ as
[\ref{Cardy87}]
\begin{equation}\label{2-242}
\left\langle{+}\right|S^z_y\left|{+}
\right\rangle\sim{n_l}^{-\beta/\nu}={n_l}^{-1/2}.
\end{equation}
Hence we may have ${J_{\rm eff}}\sim{\rm O}({n_l}^0)$.

As we mentioned in (\ref{2-40}) and (\ref{2-55}), it is known
[\ref{Alcaraz88a}] for the antiferromagnetic Heisenberg chain with the coupling
${J_{\rm eff}}$ that the energy gap behaves as
\begin{equation}\label{2-245}
\Delta E\simeq\frac{{v_{\rm s}}\pi\eta}{L}{J_{\rm eff}}
=\frac{\pi^2{J_{\rm eff}}}{2L}.
\end{equation}
Comparing this with (\ref{2-190}), we arrive at the conclusion
\begin{equation}\label{2-250}
b\equiv\lim_{x\to\infty}{\tilde{\Delta}}(x)
={v_{\rm s}}\pi\eta\frac{{J_{\rm eff}}}{J}
=\frac{\pi^2}{2}\frac{{J_{\rm eff}}}{J}.
\end{equation}

\section{Perturbational derivation of the scaling ansatz}
\setcounter{equation}{0}
\label{sec:pert}
In this section, we describe a perturbational calculation which yields the
scaling form (\ref{2-10}).
This was briefly reported in [\ref{Nishiyama95b}].

\subsection{Zeroth order of ${J_\perp}/J$}
Let us first describe the energy spectrum for ${J_\perp}/J=0$.
At this point, we have ${n_l}$ number of the $S=1/2$ Heisenberg chains
independent of each other.
The spectrum of each chain is given [\ref{Alcaraz88a}] as
figure \ref{spectrum}.
\begin{figure}[t]
\begin{minipage}[t]{99mm}
\special{epsfile=fig5_pre.eps hsize=255}
\hspace*{99mm}\vspace*{63mm}
\end{minipage}
\begin{minipage}[t]{83mm}
\caption{spectrum}%
{A schematic view of the low-lying energy spectrum of the $S=1/2$
antiferromagnetic Heisenberg chain.}
\end{minipage}
\end{figure}
Let $\left|{{\rm s}_y}\right\rangle$ denote the singlet ground state of the
$y$th leg, and let
$\left|{M_y}\right\rangle=\{\left|{1_y}\right\rangle,\left|{0_y}\right\rangle,
\left|{-1_y}\right\rangle\}$ denote the triplet of the first
excited states.
The energy of the singlet ground state is written in the form
\begin{equation}\label{3-10}
E_{\rm s}\simeq\epsilon_0LJ-\frac{{v_{\rm s}}\pi c}{6L}J
=\left(\epsilon_0L-\frac{\pi^2}{12L}\right)J
\qquad\mbox{as}\quad
L\to\infty,
\end{equation}
where $\epsilon_0=1/4-\ln2$ is the exact ground-state energy density
[\ref{Bethe31}], ${v_{\rm s}}=\pi/2$ is the sound velocity as before, and $c=1$
denotes the central charge.
The energy gap to the triplet states is given by
\begin{equation}\label{3-20}
\Delta E\simeq\frac{{v_{\rm s}}\pi\eta}{L}J=\frac{\pi^2}{2L}J
\qquad\mbox{as}\quad
L\to\infty,
\end{equation}
where $\eta=1$ is Fisher's correlation exponent as before.

The ground state of the whole system is given by the direct product of
the singlet states as
\begin{equation}\label{3-30}
\left|{\psi_{\rm gs}^{(0)}}\right\rangle
=\bigotimes_{y=1}^{n_l}\left|{{\rm s}_y}\right\rangle
\end{equation}
with the energy
\begin{equation}\label{3-40}
E_{\rm gs}\simeq{n_l}\left(\epsilon_0L-\frac{\pi^2}{12L}\right)J.
\end{equation}
The first excited states have the $3{n_l}$-fold degeneracy:
\begin{equation}\label{3-50}
\left|{\psi_{\rm ex}^{(0)}}\right\rangle
=\left|{M_{y_1}}\right\rangle\otimes
\left(\bigotimes_{{y=1}\atop{y\neq y_1}}^{n_l}
\left|{{\rm s}_y}\right\rangle\right)
\qquad
(\mbox{$M=0,\pm1$ and $y_1=1,2,\cdots,{n_l}$}).
\end{equation}
The energy gap to these excited states is the same as in (\ref{3-20}).

\subsection{First-order perturbation of ${J_\perp}/J$}
Now we calculate the first-order perturbation with respect to
${{\cal H}_{\rm rung}}$.
The first-order energy of the ground state is given by
\begin{equation}\label{3-60}
E_{\rm gs}^{(1)}=\left\langle{\psi_{\rm gs}^{(0)}}\right|
{{\cal H}_{\rm rung}}\left|{\psi_{\rm gs}^{(0)}}\right\rangle
={J_\perp}\sum_{x=0}^{L-1}\sum_{y=1}^{{n_l}-1}
\left\langle{{\rm s}_y}\right|\vec{S}_{x,y}\left|{{\rm s}_y}\right\rangle\cdot
\left\langle{{\rm s}_{y+1}}\right|\vec{S}_{x,y+1}\left|{{\rm s}_{y+1}}
\right\rangle
=0,
\end{equation}
because we have
\begin{equation}\label{3-70}
\left\langle{{\rm s}}\right|S^z_x\left|{{\rm s}}\right\rangle
=\left\langle{{\rm s}}\right|S^\pm_x\left|{{\rm s}}\right\rangle=0.
\end{equation}

In order to obtain the first-order energy of the excited state, we have to
construct the secular equation, because the zeroth-order first excited states
are degenerate.
We concentrate on the sector $\sum_yM_y=1$.
The ground state of this sector gives the perturbed first excited state,
because the perturbed energy spectrum would be similar to that in
figure \ref{spectrum} owing to the SU(2) symmetry.
The following ${n_l}$ states are of this sector:
\begin{equation}\label{3-80}
\left|{(y_1)}\right\rangle\equiv\left|{1_{y_1}}\right\rangle\otimes
\left(\bigotimes_{{y=1}\atop{y\neq y_1}}^{n_l}\left|{{\rm s}_y}\right\rangle
\right)
\qquad
(y_1=1,2,\cdots,{n_l}).
\end{equation}
We immediately have
\begin{equation}\label{3-90}
\left\langle{(y_1)}\right|{{\cal H}_{\rm rung}}\left|{(y_2)}\right\rangle=0
\qquad\mbox{unless}\quad
|y_1-y_2|=1.
\end{equation}
The nonzero matrix elements are
\begin{eqnarray}\label{3-100}
\left\langle{(y+1)}\right|{{\cal H}_{\rm rung}}\left|{(y)}\right\rangle
&=&\frac{{J_\perp}}{2}\sum_{x=0}^{L-1}
\left\langle{{\rm s}_y}\right|S^-_{x,y}\left|{1_y}\right\rangle
\left\langle{1_{y+1}}\right|S^+_{x,y+1}\left|{{\rm s}_{y+1}}\right\rangle
={J_\perp} a
\nonumber\\
&&(y=1,2,\cdots,{n_l}-1)
\end{eqnarray}
and their conjugates.
Here the coefficient $a$ is defined by
\begin{equation}\label{3-110}
a\equiv
\frac{L}{2}\left|\left\langle{{\rm s}}\right|S^-_{x}\left|{1}\right\rangle
\right|^2
\end{equation}
(The $x$ dependence of the matrix element $\left\langle{{\rm s}}
\right|S^-_{x}\left|{1}\right\rangle$ appears
only in the phase factor in the form ${\rm e}^{{\rm i}kx}$, and hence the
absolute value is independent of $x$.)
The secular equation for the first-order energy of the excited state is
thereby written in the form
\begin{equation}\label{3-120}
\left|
\begin{array}{ccccc}
-E_{\rm ex}^{(1)} & {J_\perp} a & & & {\smash{\lower1.7ex\hbox{\bg 0}}} \\
{J_\perp} a & -E_{\rm ex}^{(1)} & {J_\perp} a & & \\
& {J_\perp} a & \ddots & \ddots & \\
& & \ddots & \ddots & {J_\perp} a \\
{\smash{\hbox{\bg 0}}} & & & {J_\perp} a & -E_{\rm ex}^{(1)}
\end{array}
\right|=0.
\end{equation}
This is equivalent to the one-dimensional tight-binding model under the free
boundary condition;
the triplet state hops from a leg to a neighboring leg with the hopping
amplitude ${J_\perp} a$.
This model is exactly solvable [\ref{Matsubara73}].
The zeroth-order wave function and the first-order energy are obtained in the
forms
\begin{eqnarray}\label{3-130}
\left|{\psi_{\rm ex}^{(0)}}\right\rangle
&=&\sqrt{\frac{2}{{n_l}}}\sum_{y=1}^{n_l}
\sin\left(\frac{m\pi}{{n_l}+1}y\right)\left|{(y)}\right\rangle,
\\
\label{3-135}
E_{\rm ex}^{(1)}&=&-2{J_\perp} a\cos\frac{m\pi}{{n_l}+1},
\end{eqnarray}
for $m=1,2,\cdots,{n_l}$.
The state with $m=1$ gives the lowest energy.

We thereby obtain the energy gap up to the first order of ${J_\perp}/J$ as
\begin{equation}\label{3-140}
\Delta E\simeq\frac{J}{L}
\left(\frac{\pi^2}{2}-2xa\cos\frac{\pi}{{n_l}+1}\right),
\end{equation}
where $x\equiv L{J_\perp}/J$ as before.

We are in position to estimate the size dependence of the coefficient $a$.
In the following we show analytically and numerically that $a={\rm O}(L^0)$.

We first show that the coefficient $a$ is proportional to the transverse
susceptibility of the Heisenberg chain.
Let us consider the Hamiltonian with a transverse field at the origin:
\begin{equation}\label{3-150}
{\tilde{\Delta}}e{{\cal H}}\equiv
J\sum_{i=0}^{L-1}\vec{S}_i\cdot\vec{S}_{i+1}
-\gamma S^x_0.
\end{equation}
After the standard calculation, we have
\begin{eqnarray}\label{3-160}
\chi_{00}&\equiv&\left.\frac{\partial}{\partial \gamma}
\left\langle{S^x_0}\right\rangle_\gamma\right|_{\gamma=0}
=\sum_{\psi_n(\neq\psi_{\rm gs})}
\frac{\left|\left\langle{\psi_n}\right|S^x_0\left|{\psi_{\rm gs}}
\right\rangle\right|^2}{\Delta E_n}
\\
\label{3-170}
&=&\int_0^\infty{\rm d}\tau
\left\langle{S^x_0S^x_0(\tau)}\right\rangle_0,
\end{eqnarray}
where the angular brackets $\left\langle{\cdots}\right\rangle_\gamma$ denote
the expectation value with respect to the ground state of (\ref{3-150}),
$\Delta E_n$ is the excitation energy of the excited state
$\left|{\psi_n}\right\rangle$, and
\begin{equation}\label{3-175}
S^x_0(\tau)\equiv
{\rm e}^{-\tau{\tilde{\Delta}}e{{\cal H}}}S^x_0
{\rm e}^{\tau{\tilde{\Delta}}e{{\cal H}}}.
\end{equation}
The most dominant of the excited states $\left|{\psi_n}\right\rangle$ is
$\left|{S=1,S^z=\pm1}\right\rangle$ with the energy gap (\ref{3-20}), or
$\Delta E\sim L^{-1}$.
Hence the expression (\ref{3-160}) is approximately rewritten as
\begin{equation}\label{3-180}
\chi_{00}\simeq\frac{4a}{\pi^2}.
\end{equation}
On the other hand, a field-theoretic description [\ref{Luther75}] yields
\begin{equation}\label{3-190}
\left\langle{S^x_0S^x_0(\tau)}\right\rangle_0\sim\frac{1}{\tau}.
\end{equation}
Thus the expression (\ref{3-170}) is a dimensionless quantity:
$\chi_{00}={\rm O}(L^0)$.
We thus arrive at the conclusion $a={\rm O}(L^0)$.
There may be logarithmic corrections to this size dependence.

We also calculated the coefficient $a$ in (\ref{3-110}) numerically for
$L\leq24$;
see figure \ref{coeff}.
\begin{figure}[t]
\begin{minipage}[t]{99mm}
\special{epsfile=fig6_pre.eps hsize=255 voffset=25}
\hspace*{99mm}\vspace*{60mm}
\end{minipage}
\begin{minipage}[t]{83mm}
\caption{coeff}%
{The size dependence of the quantity $a$ appears to be of the form
$a\sim(\ln L)^\omega$.
The solid line connects the last two data points with the slope
$\omega\simeq0.68$.}
\end{minipage}
\end{figure}
The logarithmic plot reveals that the coefficient actually behaves as
\begin{equation}\label{3-200}
a\sim(\ln L)^{0.68}={\rm O}(L^0).
\end{equation}

\subsection{Order estimation of higher-order perturbations}
We next derive the approximate second-order energy of the ground state.
The second-order perturbation is given by the formula
\begin{equation}\label{3-210}
E_{\rm gs}^{(2)}=-\sum_{\psi_n(\neq\psi_{\rm gs})}
\frac{\left|\left\langle{\psi_n}\right|{{\cal H}_{\rm rung}}
\left|{\psi_{\rm gs}}\right\rangle\right|^2}{\Delta E_n}.
\end{equation}
The most dominant excited states are the $3({n_l}-1)$ states
\begin{equation}\label{3-220}
\left|{\psi_n}\right\rangle=\left\{
\begin{array}{c}
\left|{0_{y_1}0_{y_1+1}}\right\rangle\\
\left|{1_{y_1}-1_{y_1+1}}\right\rangle\\
\left|{-1_{y_1}1_{y_1+1}}\right\rangle
\end{array}
\right\}
\otimes
\left(\bigotimes_{{y=1}\atop{y\neq y_1,y_1+1}}^{{n_l}}
\left|{{\rm s}_y}\right\rangle\right)
\end{equation}
with $y_1=1,2,\cdots,{n_l}-1$.
The energy gap $\Delta E_n$ to these states is twice as large as in
(\ref{3-20}):
\begin{equation}\label{3-230}
\Delta E_n\simeq\frac{\pi^2J}{L}.
\end{equation}
The numerator of (\ref{3-210}) gives the factor $a^2$ for each of the states
(\ref{3-220}).
Hence the second-order energy of the ground state is
\begin{equation}\label{3-240}
E_{\rm gs}^{(2)}\simeq
-\frac{3({n_l}-1)a^2}{\pi^2}\left(\frac{L{J_\perp}}{J}\right)^2\frac{J}{L}.
\end{equation}

Although it is complicated to calculate the second-order energy of the first
excited state (\ref{3-130}), we can see that the size dependence is the same
as in (\ref{3-240}):
\begin{equation}\label{3-250}
E_{\rm ex}^{(2)}\sim
a^2\left(\frac{L{J_\perp}}{J}\right)^2\frac{J}{L}.
\end{equation}
We thereby have
\begin{equation}\label{3-260}
\Delta E\simeq\frac{J}{L}
\left[\frac{\pi^2}{2}+\alpha_1ax
+\alpha_2\left(ax\right)^2\right],
\end{equation}
where $x\equiv L{J_\perp}/J$, and $\alpha_1$ and $\alpha_2$ are constants.
Since $a$ is of the order of $L^0$, the expression (\ref{3-260}) is consistent
with the scaling form (\ref{2-10}).

It is generally difficult to calculate higher-order perturbation explicitly.
However, it is possible to estimate the size dependence roughly
[\ref{Hatano95a}].
The $k$th-order energy is approximately given by
\begin{eqnarray}\label{3-270}
E^{(k)}
\sim
&{\displaystyle \sum_{\{\psi\}}}&
\left\langle{\psi_{\rm gs}}\right|{{\cal H}_{\rm rung}}\left|{\psi_1}
\right\rangle
\left\langle{\psi_2}\right|{{\cal H}_{\rm rung}}\left|{\psi_3}\right\rangle
\cdots
\left\langle{\psi_{k-1}}\right|{{\cal H}_{\rm rung}}\left|{\psi_{\rm gs}}
\right\rangle
\nonumber\\
&&
\times
\left[
\left(\Delta E_1\right)
\left(\Delta E_2\right)
\cdots
\left(\Delta E_{k-1}\right)
\right]^{-1}.
\end{eqnarray}
We may estimate the dimensionality of the operator ${{\cal H}_{\rm rung}}$ at
$L{J_\perp}\times[\vec{S}]^2\sim{J_\perp}$, because the magnetic operator
$\vec{S}$
of the antiferromagnetic Heisenberg chain has the dimensionality
$[\vec{S}]=L^{-\beta/\nu}=L^{-1/2}$ [\ref{Cardy87}].
(This rough estimate is consistent with the estimation $a={\rm O}(L^0)$.)
On the other hand, each energy denominator $\Delta E$ is of the order of $J/L$,
because the Heisenberg chain is critical.
We thereby have the rough estimate
\begin{equation}\label{2-280}
E^{(k)}\sim{J_\perp}^k\left(\frac{J}{L}\right)^{-(k-1)}
=\frac{J}{L}\left(\frac{L{J_\perp}}{J}\right)^k.
\end{equation}
Hence the energy gap may be given by the form
\begin{equation}\label{2-285}
\Delta E\sim\frac{J}{L}\sum_k\alpha_kx^k
\end{equation}
with appropriate coefficients $\{\alpha_k\}$.
This is consistent with the scaling ansatz (\ref{2-10}).

We have not confirmed that the above perturbational expansion is
convergent.
However, the numerical results in figure \ref{scale} do suggest that the
series converges at least over a finite region of $x=L{J_\perp}/J$.

\section{Summary}
\setcounter{equation}{0}
\label{sec:summary}
In the present paper, we introduced the finite-size scaling form of the energy
gap of the antiferromagnetic Heisenberg ladder models:
\begin{equation}\label{4-10}
\Delta E(L)/J\simeq L^{-1}{\tilde{\Delta}}(L{J_\perp}/J).
\end{equation}
We confirmed this scaling form numerically as well as by means of perturbation
theory.
On the basis of the scaling theory,
we discussed the criticality of the ladder models in a unified way.
The difference between the ladders with even ${n_l}$ and with odd ${n_l}$ was
attributed to the different asymptotic behavior of the scaling function in the
limit $L{J_\perp}/J\to\infty$.

For even ${n_l}$, the energy gap develops in the form $\Delta E\sim{J_\perp}$.
The whole region of ${J_\perp}>0$ is a disordered phase with a unique ground
state.
(This ground state is reduced to a set of independent singlets in the limit
${J_\perp}/J\to\infty$.)
We estimated for ${n_l}=2$ the critical amplitude of the energy gap and the
correlation length around the critical point ${J_\perp}/J=0$.

For odd ${n_l}$, on the other hand, the whole region of ${J_\perp}>0$ is the
critical line.
The energy gap vanishes in the form $\Delta E\sim L^{-1}$ for any ${J_\perp}$.
The critical line is controlled by the $S=1/2$ antiferromagnetic Heisenberg
chain which is obtained in the limit ${J_\perp}/J\to\infty$.
The central charge of the phase is hence unity.

\section*{Acknowledgments}
The authors are grateful to Dr A Terai for helpful discussions.
The numerical calculations were performed on the super-computer HITAC S3800/480
of the Computer Centre, University of Tokyo and on the work-station HP Apollo
9000/735 of the Suzuki group, Department of Physics, University of Tokyo.

\section*{References}

\newcounter{bibnumcounter}
\newenvironment{bibnumlist}[1]{
  \begin{list}{[\arabic{bibnumcounter}]}
    {\usecounter{bibnumcounter}
     \settowidth{\labelwidth}{[#1]}
     \setlength{\leftmargin}{1.2\labelwidth}
     \setlength{\labelsep}{0.2\labelwidth}
     \setlength{\rightmargin}{0mm}
     \setlength{\parsep}{0mm}
     \setlength{\itemsep}{0mm}
    }
}{
  \end{list}
}

\begin{bibnumlist}{99}

\item\label{Cardy84a}
{Cardy J L 1984
{\em J.\ Phys.\ A: Math.\ Gen.}\ {\bf 17} L385-7}

\item\label{Cardy87}
{Cardy J L 1987
{\em Phase Transitions and Critical Phenomena} vol 11 ed C Domb and J L
Lebowitz (New York: Academic) pp 55-126}

\item\label{Takano92}
{Takano M, Hiroi Z, Azuma M and Takeda Y 1992
{\em Mechanisms of Superconductivity}
JJAP Ser.\ 7 (Tokyo: JJAP) p 3}

\item\label{White94}
{White S R, Noack R M and Scalapino D J 1994
{\em Phys.\ Rev.\ Lett.}\ {\bf 73} 886-9}

\item\label{Bethe31}
{Bethe H A 1931
{\em Z.\ Phys.}\ {\bf 71} 205-26}

\item\label{Dagotto92}
{Dagotto E, Riera J and Scalapino D 1992
{\em Phys.\ Rev.\ B} {\bf 45} 5744-7}

\item\label{Barnes93}
{Barnes T, Dagotto E, Riera J and Swanson E S 1993
{\em Phys.\ Rev.\ B} {\bf 47} 3196-203}

\item\label{Nishiyama95b}
{Nishiyama Y, Hatano N and Suzuki M 1995
{\em J.\ Phys.\ Soc.\ Japan} {\bf 64} No.\ 6}

\item\label{Terai95}
{Terai A 1995
private communication}

\item\label{Lanczos50}
{Lanczos C 1950
{\em J.\ Res.\ NBS} {\bf 45} 255-82}

\item\label{Reigrotzki95}
{Reigrotzki M, Tsunetsugu H and Rice T M 1994
{\em J.\ Phys.: Condens.\ Matter} {\bf 6} 9235-45}

\item\label{Hatano94c}
{Hatano N, Nishiyama Y and Suzuki M 1994
{\em J.\ Phys.\ A: Math.\ Gen.}\ {\bf 27} 6077-89}

\item\label{Fisher64b}
{Fisher M E 1964
{\em J.\ Math.\ Phys.}\ {\bf 5} 944-62}

\item\label{Alcaraz88a}
{Alcaraz F C, Barber M N and Batchelor M T 1988
{\em Ann.\ Phys.\ (N.\ Y.)} {\bf 182} 280-343}

\item\label{Matsubara73}
{Matsubara F and Katsura S 1973
{\em Prog.\ Theor.\ Phys.}\ {\bf 49} 367-8}

\item\label{Luther75}
{Luther A and Peschel I 1975
{\em Phys.\ Rev.\ B} {\bf 12} 3908-17}

\item\label{Hatano95a}
{Hatano N 1995
{\em J.\ Phys.\ Soc.\ Japan} {\bf 64} 1529-51}

\end{bibnumlist}
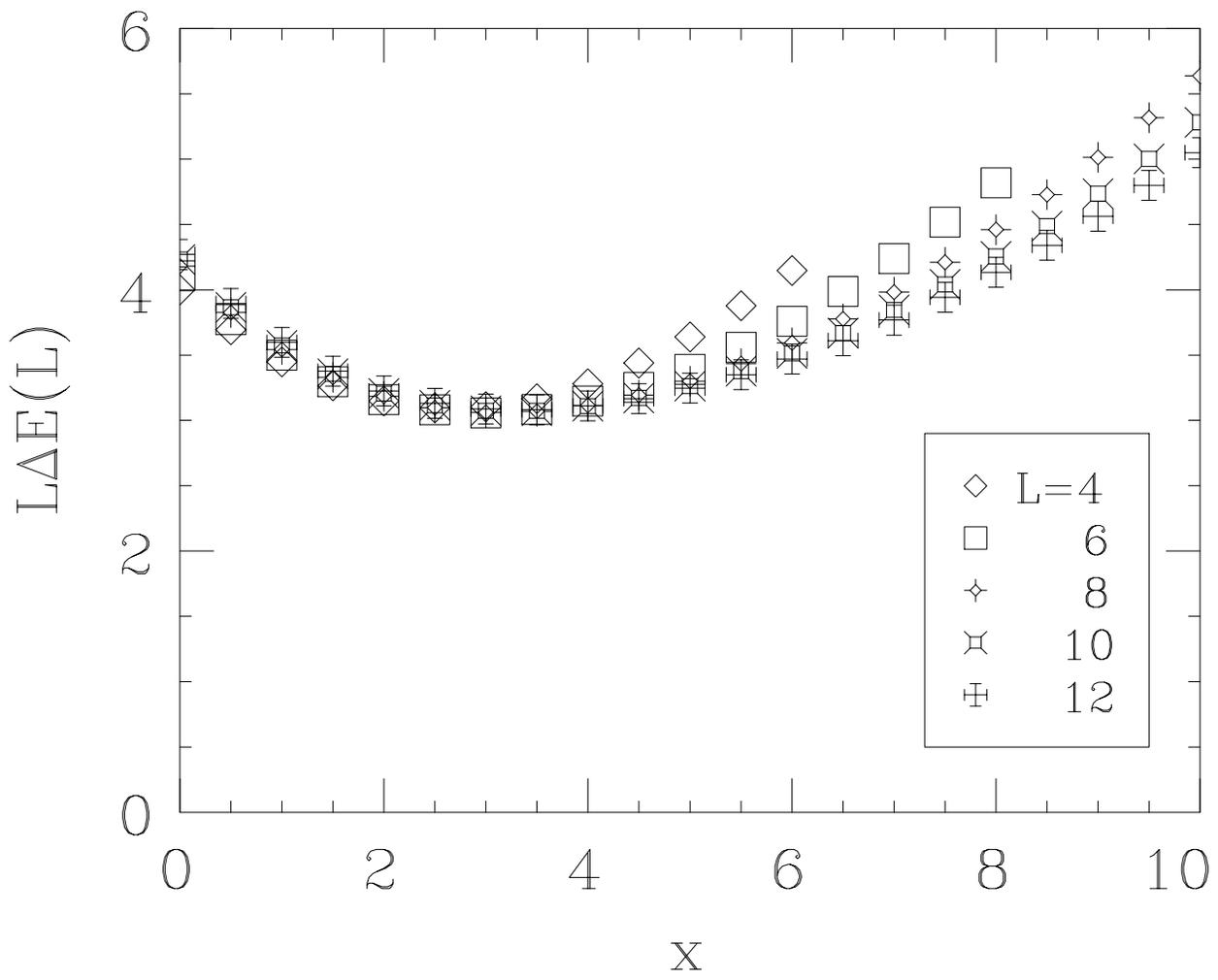
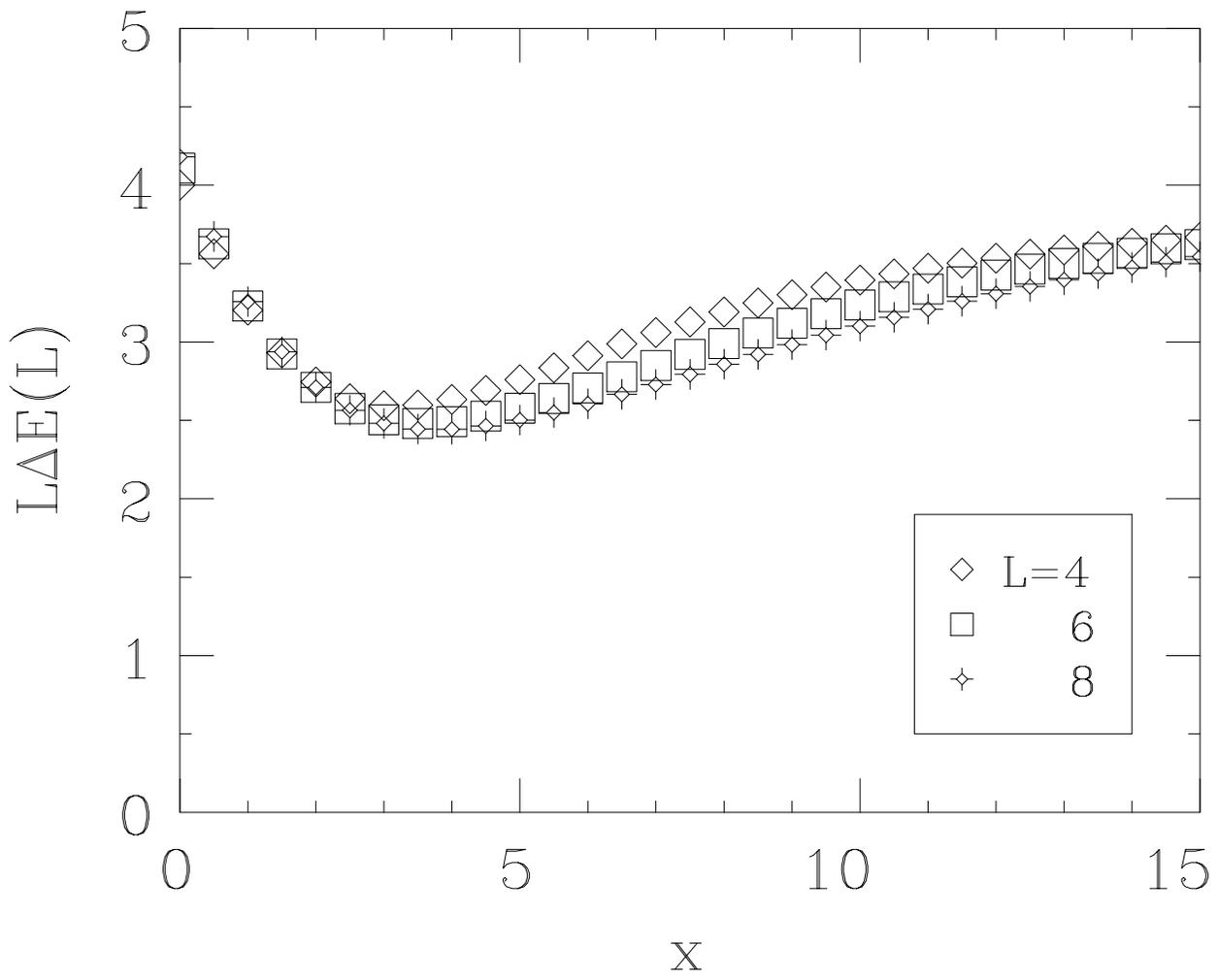
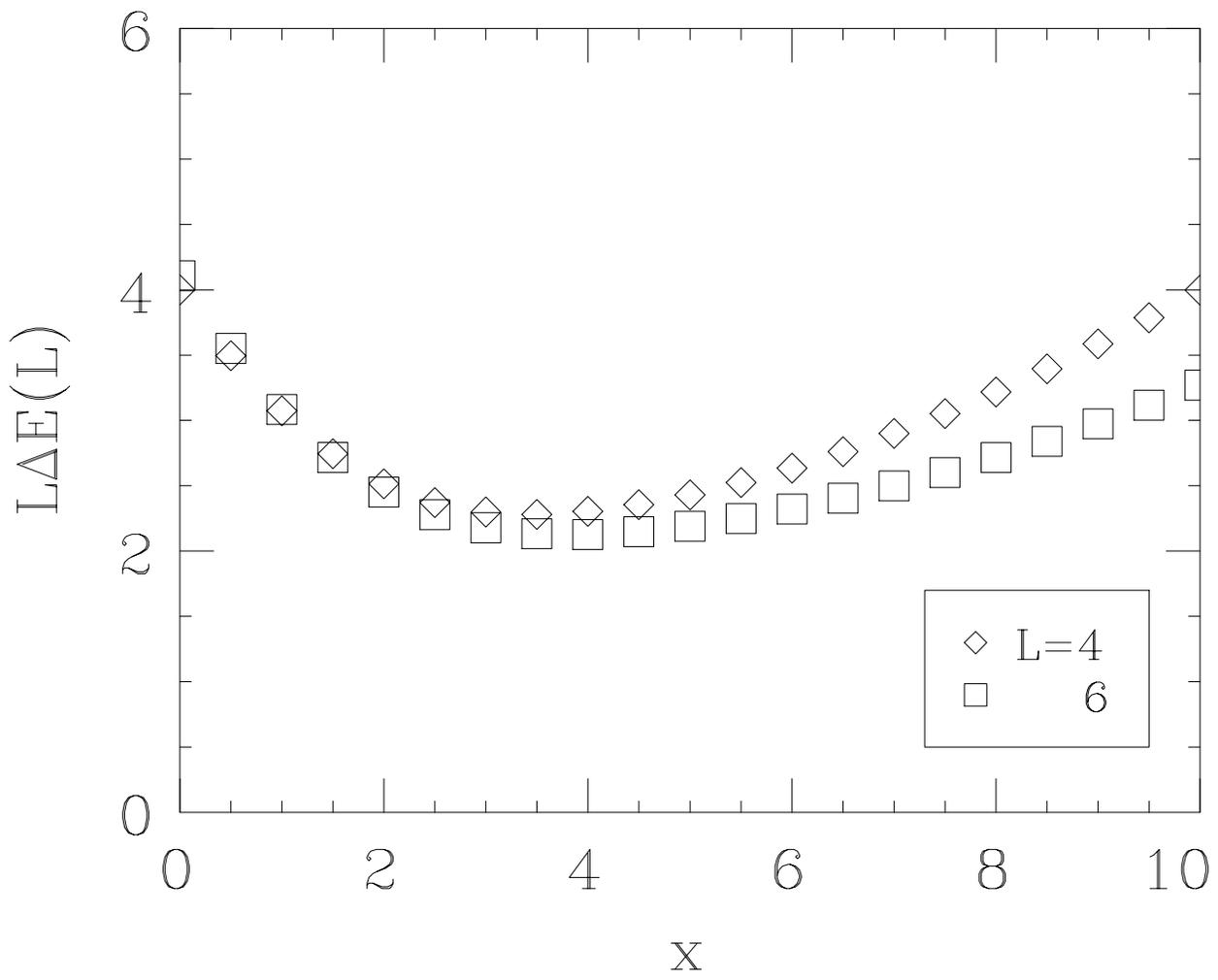
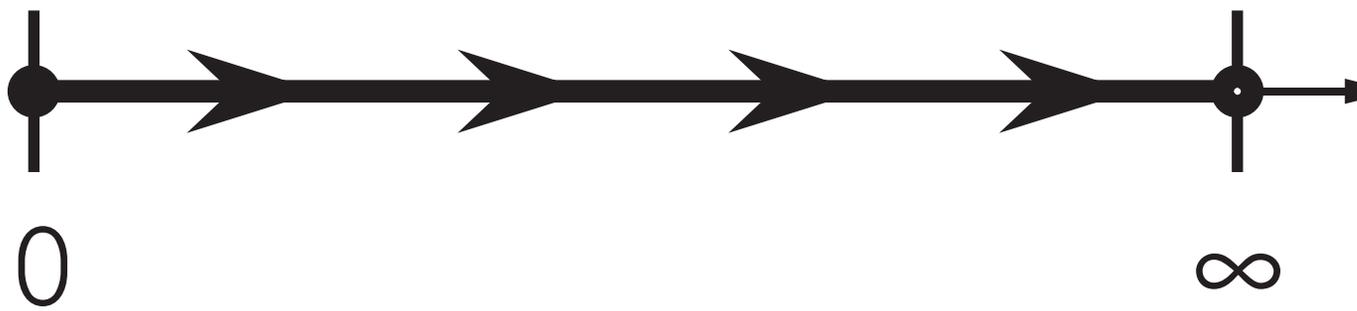
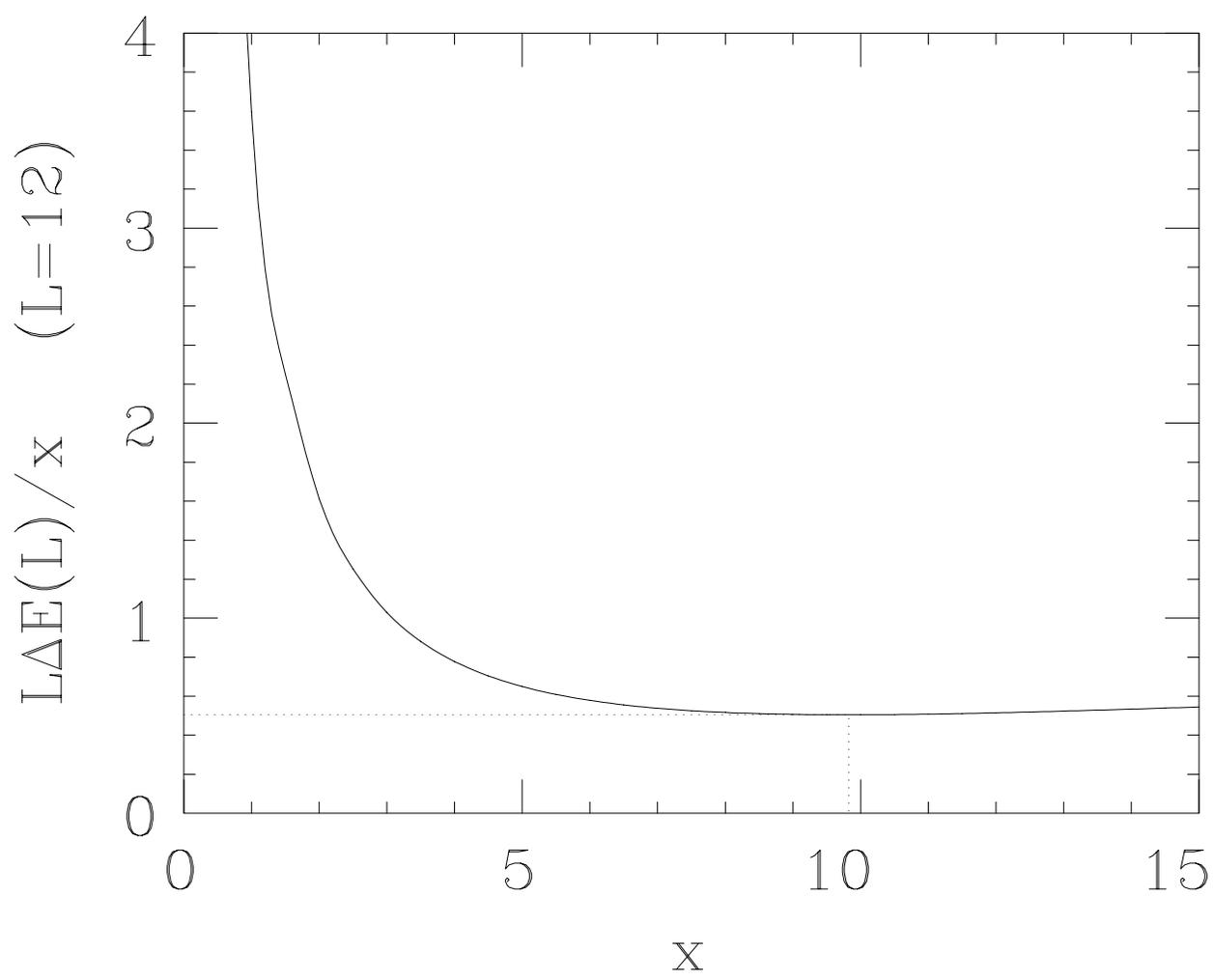
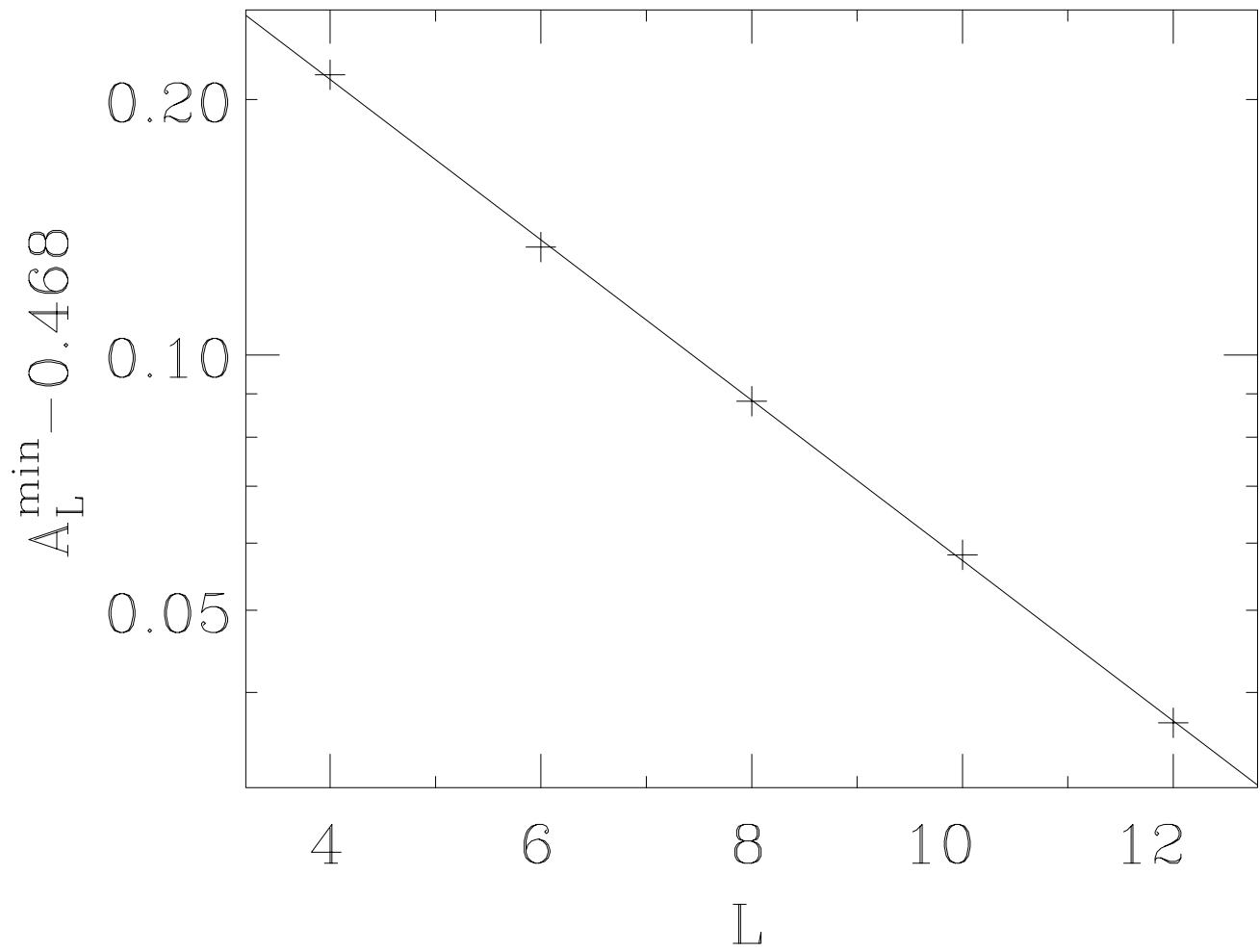
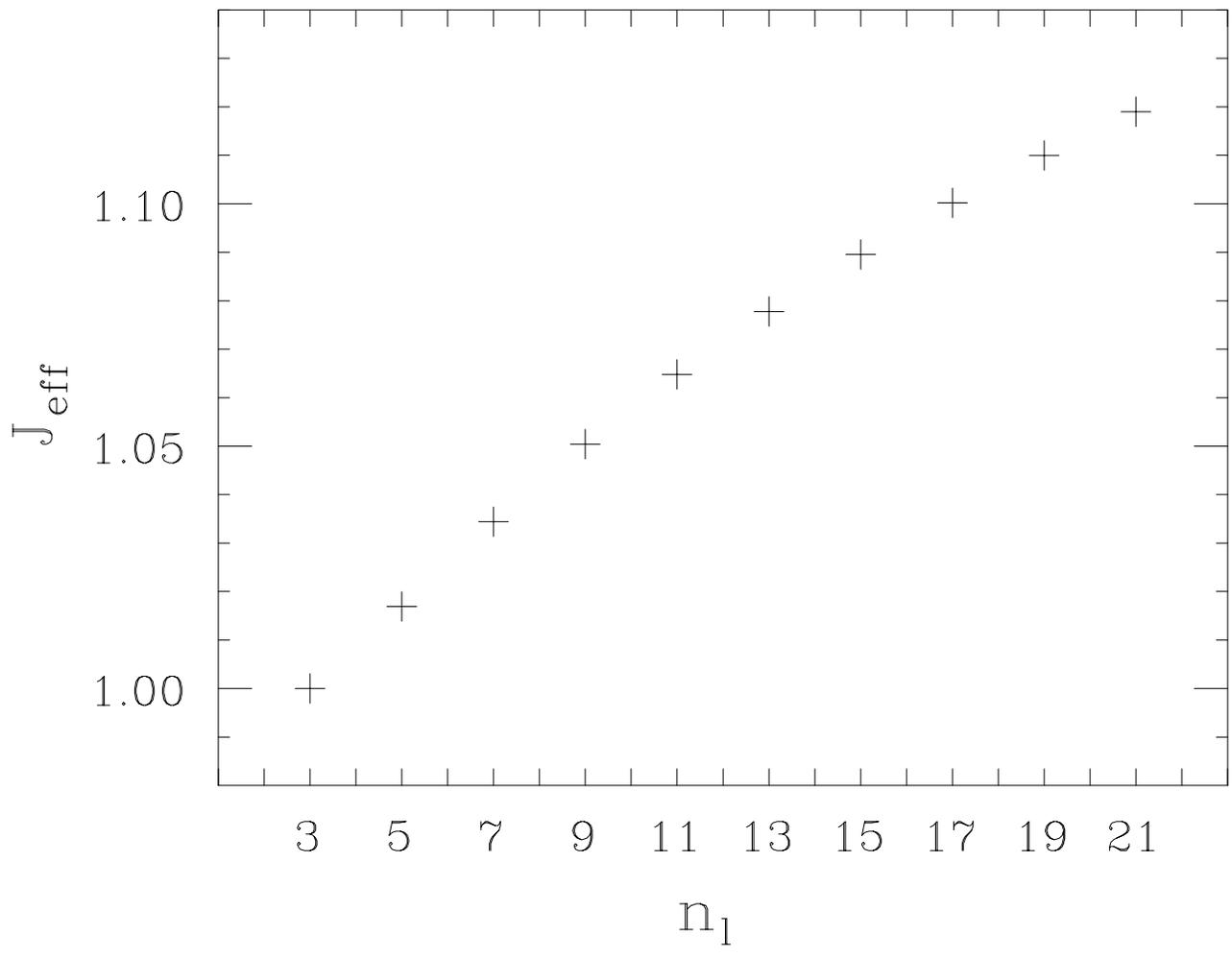
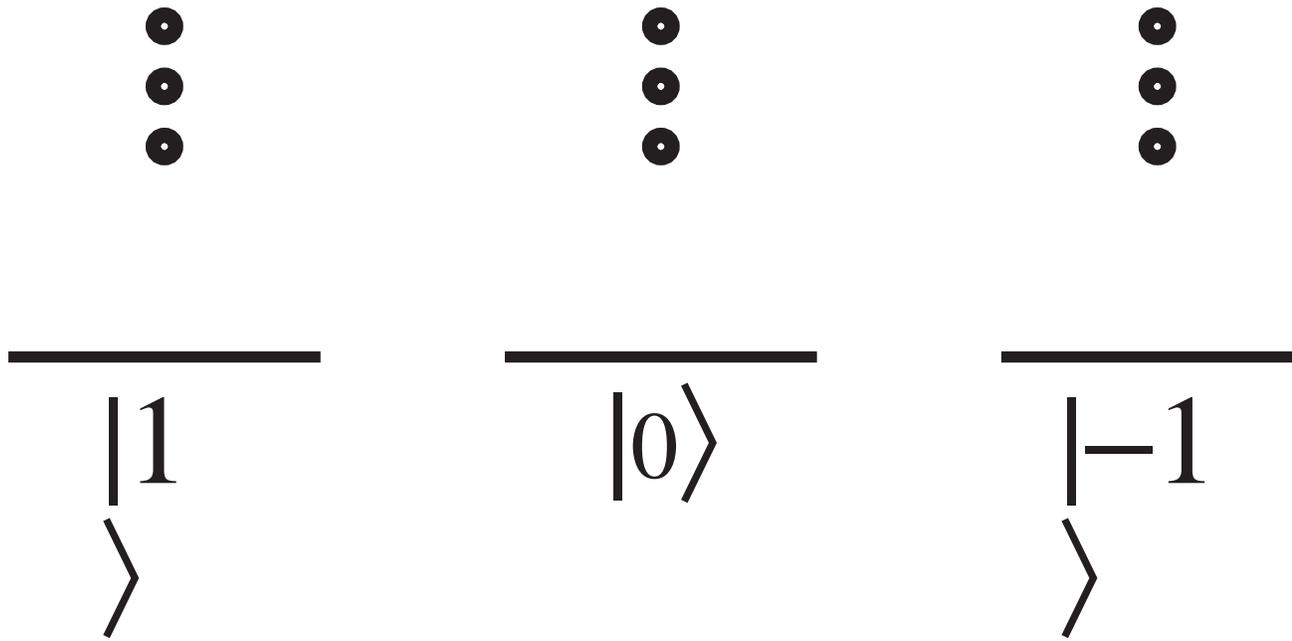
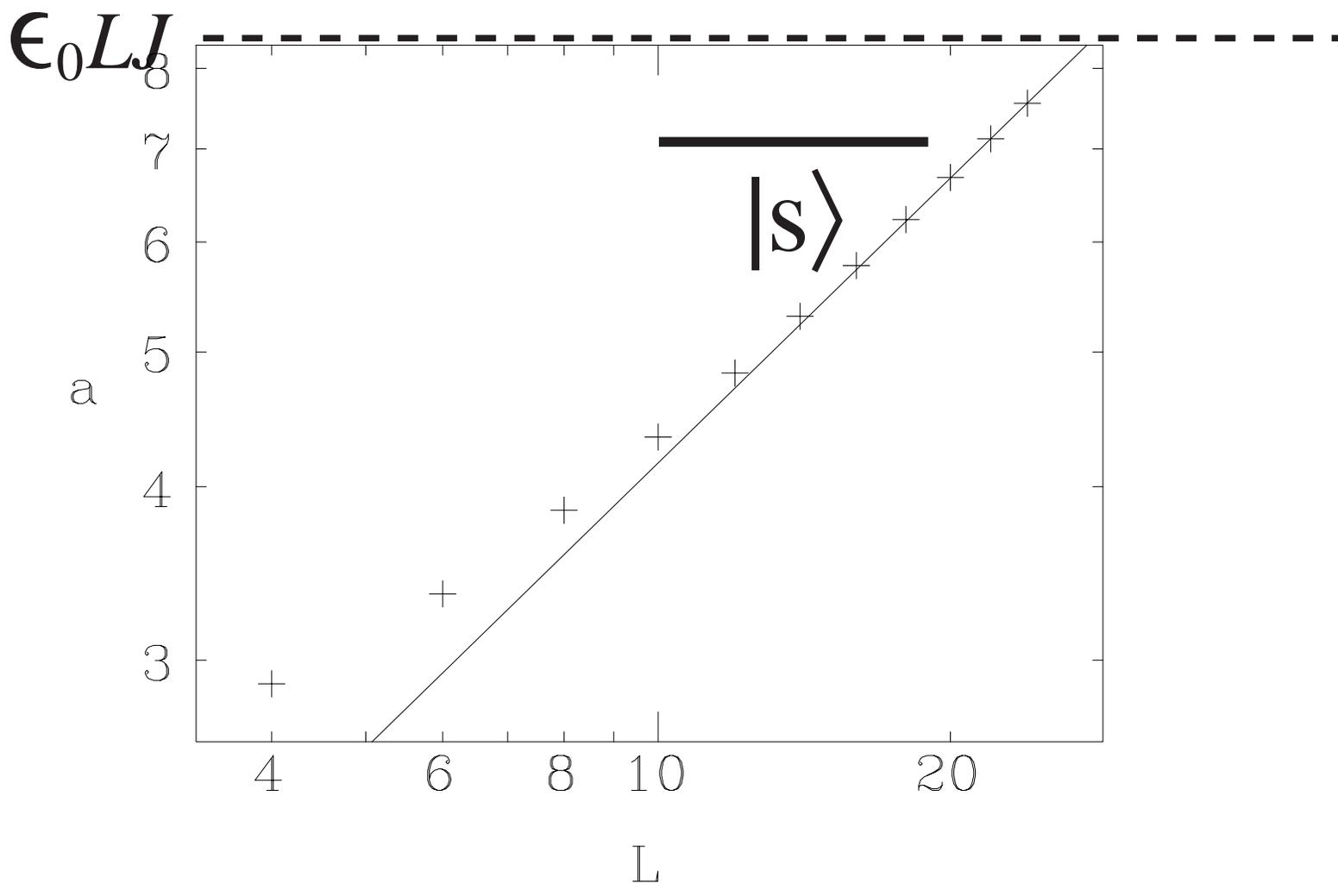
\end{document}